\documentclass[graphicx]{JHEP}
\usepackage{epsfig,amssymb}

\def\be{\begin{equation}}
\def\ee{\end{equation}}
\def\bea{\begin{array}}
\def\eea{\end{array}}
\def\beqa{\begin{eqnarray}}
\def\eeqa{\end{eqnarray}}
\def\beqas{\begin{eqnarray*}}
\def\eeqas{\end{eqnarray*}}

\def\bp{\begin{picture}}
\def\ep{\end{picture}}
\def\bc{\begin{center}}
\def\ec{\end{center}}
\def\bfig{\begin{figure}}
\def\efig{\end{figure}}

\def\bit{\begin{itemize}}
\def\eit{\end{itemize}}
\def\nn{\nonumber}
\def\f{\frac}

\def\[{\left[}
\def\]{\right]}
\def\({\left(}
\def\){\right)}

\def\..{\left.}
\def\.{\right.}
\def\tl{\tilde}

\def\la{\leftarrow}

\def\tm{\times}

\def\da{\dagger}

\def\la{\lambda}

\def\al{\alpha}

\def\ep{\epsilon}
\def\rh{\rho}

\def\ga{\gamma}
\def\pa{\partial}
\def\pr{\prime}
\def\eqv{\equiv}

\title{Realistic Flipped SU(5) from Orbifold SO(10)}
\author{Csaba Balazs$^1$, Zhaofeng Kang$^2$, Tianjun Li$^{2,3}$, Fei
Wang$^1$, Jin Min Yang$^2$\\
$^1$ School of Physics, Monash University, Melbourne Victoria 3800,
Australia\\
$^2$ Key Laboratory of Frontiers in Theoretical Physics, Institute
of Theoretical Physics, Chinese Academy of Sciences,
Beijing 100190, P. R. China \\
$^3$ George P. and Cynthia W. Mitchell Institute for Fundamental
Physics, Texas A$\&$M University, College Station, TX 77843, USA }

\abstract{ We propose a realistic flipped $SU(5)$ model derived from
a five-dimensional orbifold $SO(10)$ model. The Standard Model (SM)
fermion masses and mixings are explained by combining the
traditional Froggatt-Nielsen mechanism with the five-dimensional
wave function profiles of the SM fermions. Employing tree-level
spontaneous R-symmetry breaking in the hidden sector and
extra(ordinary) gauge mediation, we obtain realistic supersymmetry
breaking soft mass terms with non-vanishing gaugino masses.
Including the messenger fields at the intermediate scale and
Kaluza-Klein states at the compactification scale, we study gauge
coupling unification. We show that the $SO(10)$ unified gauge
coupling is very strong and the unification scale can be much higher
than the compactification scale. We briefly discuss proton decay as
well.

}

\begin{document}
\maketitle \indent
\newpage
\section{Introduction}
As one of the most attractive extensions of the Standard Model (SM),
supersymmetric Grand Unification Theories (GUTs) like
$SU(5)$~\cite{su5} or $SO(10)$~\cite{so10}
give us deep insights into the problems such as charge quantization,
neutrino masses and mixings as well as the origin of the Yukawa sector.
However, these theories still have some unsatisfactory features such as
the doublets-triplet (D-T) splitting problem, rapid proton decay, and
unrealistic SM fermion mass relations, etc.

The $SO(10)$ models are pretty interesting since they have both
the gauge interaction unification and the SM fermion unification.
One type of these models, where the gauge symmetry is broken down
to the Georgi-Glashow $SU(5)$, have the rapid proton decay and D-T
splitting problems in the subsequent gauge symmetry breaking into
the SM. In contrast, another type with symmetry breaking to
flipped $SU(5)$ might be more attractive because the flipped
$SU(5)$ models can solve the D-T splitting problem via missing
partner mechanism as well as the dimension-five proton decay
problem~\cite{f51,f52,f53}. Although embedding flipped $SU(5)$
into $SO(10)$ can retrieve gauge unification, the missing partner
mechanism does not work in four-dimensional models (for a possible
solution, see Ref.~\cite{Huang:2006nu}). A simple solution is to
realize such embedding in five-dimensional orbifold. Orbifold GUT
models for $SU(5)$ were proposed in~\cite{Kawamura:1999nj,
Kawamura:2000ev,Kawamura:2000ir} and widely studied thereafter
in~\cite{at,Hall:2001pg,Kobakhidze:2001yk,Hebecker:2001wq,
Hebecker:2001jb,Li:2001qs, Li:2001wz,fei1,fei2}. Orbifold $SO(10)$
models with symmetry breaking to Pati-Salam models were studied
in~\cite{dm, kr}. And earlier studies on orbifold $SO(10)$ models
with symmetry breaking to flipped $SU(5)$ can be found
in~\cite{barrd,ilja}.

In this paper we consider a realistic flipped $SU(5)$ model derived
from the five-dimensional orbifold $SO(10)$ and study its phenomenological
consequence. As we know, it is interesting to explain the SM
fermion masses and mixings in the Minimal Supersymmetric Standard
Model (MSSM) from the top-down approach. In particular, the
Froggat-Nielson mechanism~\cite{FN} can be very predictive in the
GUTs. The efforts to explain the flavor structure through the deformed
Froggat-Nielson mechanism in orbifold $SU(5)$ models were shown
in~\cite{KJH,YM,YMS}, in which the SM fermion mass and mixing
hierarchies are obtained via wave-function profiles of the SM
fermions by adding bulk mass terms~\cite{AJ}. However, we find that
in the flipped $SU(5)$ model it is not as simple as in the ordinary
$SU(5)$ model to explain the SM fermion masses and mixings by such
Froggat-Nielson mechanism because of the flipping of the right-handed
up- and down-type quarks. Besides, the neutrino masses and mixings
obtained from double see-saw mechanism set stringent constraints on
the possible quark mass hierarchies in the flipped $SU(5)$ model.
Therefore, we will introduce an additional discrete $Z_3$ symmetry, and
combine the traditional Froggat-Nielsen mechanism with the
wave-function profiles of the SM fermions. In this way
we can generate the observed SM fermion masses and mixings.

In addition, we will discuss the relevant problems on supersymmetry (SUSY)
breaking. We use the tree-level spontaneously R-symmetry
breaking model and (extra)ordinary gauge mediation to obtain the
realistic SUSY breaking soft mass terms with non-vanishing gaugino
masses, which is in contrast to the previous models with vanishing gaugino
masses by direct gauge mediation. Moreover, by including the messenger
fields at the intermediate scale and the Kaluza-Klein (KK) states at the
compactification scale, we will study the gauge coupling unification in
details. Our study shows that the $SO(10)$ unified gauge coupling is very
strong, and the unification scale can be much higher than the
compactification scale. We will also comment on the proton decay.

This paper is organized as follows. In Section~\ref{sec-1} we
recapitulate the flipped $SU(5)$ model. In Section~\ref{sec-2} we
present the orbifold $SO(10)$ models where the gauge symmetry is
broken down to the flipped $SU(5)$. In Section~\ref{sec-2P} we
explain the SM fermion masses and mixings via the usual
Froggat-Nielson mechanism and wave function profiles of the SM
fermions. In Section~\ref{sec-3} we discuss the four-dimensional
$N=1$ supersymmetry breaking via the tree-level spontaneously
R-symmetry breaking in hidden sector and (extra)ordinary gauge
mediation. In Section~\ref{sec-4} we discuss the strongly coupled
gauge coupling unification with the threshold corrections from the
messenger fields and KK states. In Section~\ref{sec-5} we discuss
the proton decay problem. Section~\ref{sec-6} contains our
conclusions.

\section{Flipped $SU(5)$ model }
\label{sec-1}
In this section we briefly review the
four-dimensional flipped $SU(5)$ model~\cite{f51,f52,f53}. The gauge
group for the flipped $SU(5)$ model is $SU(5)\times U(1)_{X}$, which
can be embedded in the $SO(10)$ group. We define the generator
$U(1)_{Y'}$ in $SU(5)$ as
\begin{eqnarray}
T_{\rm U(1)_{Y'}} \equiv {\rm diag} \left(-{1\over 3}, -{1\over 3},
-{1\over 3}, {1\over 2}, {1\over 2} \right).
\end{eqnarray}
The hypercharge is given by
\begin{eqnarray} Q_{Y} =
{1\over 5} \left( Q_{X}-Q_{Y'} \right).
\end{eqnarray}
The SM fermions transform under $SU(5)\times U(1)_{X}$ as follows
\begin{eqnarray}
F_i={\bf (10,1)},~ {\bar f}_i={\bf (\bar{5},-3)},~l_i^c={\bf (1,
5)}, \label{smfermions}
\end{eqnarray}
where $i=1, 2, 3$. And the particle assignments are
\begin{eqnarray}
F_i=(Q_i, D^c_i, N^c_i),~{\bar f}_i=(U^c_i, L_i),~l_i^c=E^c_i.
\end{eqnarray}
where $Q_i$ and $L_i$ are the quark and lepton doublet superfields,
and $U^c_i$, $D^c_i$, $E^c_i$, and $N^c_i$ are the charge conjugate
superfields of the right-handed up-type quark, down-type quark,
lepton and neutrino, respectively.

To break the GUT and electroweak gauge symmetries, two pairs of
Higgses are introduced in the following representations
\begin{eqnarray}
H={\bf (10, 1)},~~{\overline H}={\bf ({\overline{10}}, -1)},
~~h={\bf (5, -2)},~~{\bar h}={\bf ({\bar {5}}, +2)}~.
\label{Higgse1}
\end{eqnarray}
We label the states in the Higgs multiplets by the same symbols as
in the SM fermion multiplets. Explicitly, the Higgs particles are
\beqa
&&H=(Q_H, D_H^c,{N}^c_H), ~~~~
 {\overline H}= ({\overline {Q}}_{\overline H}, {\overline {D}}_{\overline{H}}^c,
{\overline{N}^c}_{\overline H}), \label{Higgse2} \\
&& h=(D_h, D_h, D_h, H_d), ~~~
{\bar h}=({\overline {D}}_{\bar h}, {\overline {D}}_{\bar h},
{\overline {D}}_{\bar h}, H_u).
\label{Higgse3}
\eeqa
where $H_d$ and $H_u$ are the two Higgs doublets in the MSSM.

The flipped $SU(5)$ model elegantly solves the D-T splitting problem
via the missing partner mechanism. After $N_H^c$ and
$\overline{N}_{\overline{H}}^c$ acquire vacuum expectation values
(VEV) which break the flipped $SU(5)$ gauge symmetry down to the SM
gauge symmetry, the superfields $H$ and $\overline{H}$ will be eaten
by supersymmetric Higgs mechanism except the $D_H^c$ and
${\overline{D}}_{\overline H}^c$ components. The superpotential term
\begin{eqnarray}\label{DT}
W_{D-T}=~4\pi \left(\lambda H H h + {\bar \lambda} {\overline H}
{\overline H} {\overline h} \right)
\end{eqnarray}
couple $D_H^c$ and ${\overline {D}}^c_{\overline H}$ with respectively
$D_h$ and ${\overline {D}}_{\overline h}$ to form
heavy eigenstates with masses $8 \pi \lambda <N_H^c>$ and $8 \pi
{\bar \lambda} <{\overline {N}}^c_{\overline H}>$. But the Higgs
doublets remain massless since they do not have vector-like partners
in $H$ and $\overline{H}$. Thus, the doublets and triplets in $h$
and $\bar{h}$ are split. Because the triplets in $h$ and $\bar{h}$
only have small mixing through the effective $\mu$-term, the
Higgsino-exchange mediated proton decay are negligible, {\it i.e.},
we do not have the dimension-five proton decay problem.

\section{Flipped $SU(5)$ from Five-Dimensional
Orbifold $SO(10) $} \label{sec-2}

We consider the five-dimensional space-time ${\cal M}_4{\tm}
S^1/(Z_2{\tm}Z_2)$ comprising of the Minkowski space ${\cal M}_4$
with coordinates $x_{\mu}$ and the orbifold $S^1/(Z_2{\tm}Z_2)$
with coordinate $y{\eqv}x_5$. The orbifold $S^1/(Z_2{\tm}Z_2)$ is
obtained from $S^1$ by moduling the equivalent classes
 \beqa
 P:~y{\sim} -y~,~~~~~~~~~P^{\pr}:~y^{\pr}\sim -y^{\pr}~,
 \eeqa
where $y^{\pr}{\eqv}y+\pi R/2$. There are two inequivalent
3-branes locating at $y=0$ and $y=\pi R/2$ which are denoted as
$O$ and $O^{\pr}$, respectively.

The five-dimensional $N=1$ supersymmetric gauge theory has 8 real
supercharges, corresponding to $N=2$ supersymmetry in four
dimensions. The vector multiplet physically contains a vector boson
$A_M$ where $M=0, 1, 2, 3, 5$, two Weyl gauginos $\lambda_{1,2}$,
and a real scalar $\sigma$. In terms of four-dimensional $N=1$
language, it contains a vector multiplet $V(A_{\mu}, \lambda_1)$ and
a chiral multiplet $\Sigma((\sigma+iA_5)/\sqrt 2, \lambda_2)$ which
transform in the adjoint representation of the gauge group. And the
five-dimensional hypermultiplet physically has two complex scalars
$\phi$ and $\phi^c$, a Dirac fermion $\Psi$, and can be decomposed
into two 4-dimensional chiral mupltiplets $\Phi(\phi, \psi \equiv
\Psi_R)$ and $\Phi^c(\phi^c, \psi^c \equiv \Psi_L)$, which transform
as conjugate representations of each other under the gauge group.

The general action for the gauge fields and their couplings to the
bulk hypermultiplet $\Phi$ is~\cite{nima,nima2}
\begin{eqnarray}
S&=&\int{d^5x}\frac{1}{k g^2} {\rm
Tr}\left[\frac{1}{4}\int{d^2\theta} \left(W^\alpha W_\alpha+{\rm H.
C.}\right) \right.\nonumber\\&&\left.
+\int{d^4\theta}\left((\sqrt{2}\partial_5+ {\bar \Sigma })
e^{-V}(-\sqrt{2}\partial_5+\Sigma )e^V+
\partial_5 e^{-V}\partial_5 e^V\right)\right]
\nonumber\\&& +\int{d^5x} \left[ \int{d^4\theta} \left( {\Phi}^c e^V
{\bar \Phi}^c + {\bar \Phi} e^{-V} \Phi \right)
\right.\nonumber\\&&\left. + \int{d^2\theta} \left( {\Phi}^c
(\partial_5 -{1\over {\sqrt 2}} \Sigma) \Phi + {\rm H. C.}
\right)\right]~.~\, \label{VD-Lagrangian}
\end{eqnarray}
Possible kink mass terms can be added to hypermultiplets which will
play a central role in reproducing the SM fermion masses and mixings
in our paper.

We consider the flipped $SU(5)$ gauge theory obtained from bulk
$SO(10)$ gauge theory via orbifolding in the five-dimensional
$Z_2\times Z^\pr_2$ orbifold. We can choose proper boundary
conditions to break $SO(10)$ gauge symmetry down to flipped $SU(5)$
in the $O^\pr$ brane at $y=\pi R/2$. The boundary conditions
(($Z_2$, $Z^\pr_2$) parities) for the bulk fields can be chosen so
that the $SO(10)$ representation can be decomposed in terms of
flipped $SU(5)$
\beqa V^g({\bf 45}) &=& V^{++}_{{\bf 24}^0} +
V^{++}_{{\bf 1}^0} + V^{+-}_{{\bf
10}^{-4}} + V^{+-}_{\overline{{\bf 10}}^4} \nn\\
\Sigma^g({\bf 45})&=&\Sigma^{--}_{{\bf 24}^0} + \Sigma^{--}_{{\bf
1}^0} +\Sigma^{-+}_{{\bf 10}^{-4}} + \Sigma^{-+}_{\overline{{\bf
10}}^4},\nn\\
\Phi({\bf 16})_1 &=& \Phi^{++}_{{\bf 10}^1} +
\Phi^{+-}_{\overline{{\bf 5}}^{-3}} + \Phi^{+-}_{{\bf 1}^5},\nn\\
\Phi({\bf 16})_2 &=& \Phi^{+-}_{{\bf 10}^1} +
\Phi^{++}_{\overline{{\bf 5}}^{-3}} + \Phi^{+-}_{{\bf 1}^5},\nn\\
\Phi({\bf 16})_3 &=& \Phi^{+-}_{{\bf 10}^1} +
\Phi^{+-}_{\overline{{\bf 5}}^{-3}} + \Phi^{++}_{{\bf 1}^5},\nn\\
H({\bf 10})_1 &=& H^{++}_{{\bf 5}^{-2}} + H^{+-}_{\overline{\bf
5}^{2}},\nn\\ H({\bf 10})_2 &=& H^{+-}_{{\bf 5}^{-2}} +
H^{++}_{\overline{\bf 5}^{2}}~.
\eeqa
Also, the ($Z_2$, $Z^\pr_2$)
parities for $\Phi^c$ and $H^c$ are opposite to these of $\Phi$ and
$H$. In order to explain the SM fermion masses and mixings, we
choose the boundary conditions for ${\bf 16}$ so that we have three
types of wave function profiles for ${\bf 10}_{1}$, $\bar{\bf
5}_{-3}$, and ${\bf 1}_{5}$, respectively. This is different from
the naive orbifold $SO(10)$ models. Such boundary conditions are
possible by introducing large brane mass terms for relevant fields
to change Neumann boundary conditions into Dirichlet boundary
conditions~\cite{boundary}.

\section{The SM Fermion Masses and Mixings}
\label{sec-2P}

It is well known that the SM fermion masses and mixings exhibit a
hierarchical structure. The quark CKM mixings can be cast, in the
Wolfenstein formalism, as~\cite{CKM}
\beqa
V_{CKM}=\(\bea{ccc}1-\f{\la^2}{2}&\la& A\la^3(\rh+i\eta)\\
-\la&1-\f{\la^2}{2}&A\la^2\\A\la^3(1-\rh+i\eta)&~-A\la^2&1\eea\)~,
\eeqa where $A$ is of order 1 while $\rh$ and $\eta$ are between
$\la$ and 1. The hierarchy is reflected in the dependence of
various entries on different powers of $\la\approx 0.22$.
Renormalization group evolution (RGE) of the charged fermion
masses to a high scale ($\sim 10^{16}$ GeV) also reveals the
following hierarchical structure
\beqa
m_t~:~m_c~:~m_u~&\simeq&
~1~:~\la^4~:~\la^8~,\nn\\~m_b~:~m_s~:~m_d~&\simeq&
~1~:~\la^2~:~\la^4~,\nn\\~m_\tau~:~m_\mu~:~m_e~&\simeq&
~1~:~\la^2~:~\la^4~,
\eeqa
with $m_b/m_t=\la^3$. In this section
we discuss the explanation of the pattern of the SM fermion masses
and mixings in the flipped $SU(5)$ model.

In extra dimensional models, a well known approach to generate the
SM fermion hierarchies is the so called zero mode wave function
profile~\cite{AJ}. A non-trivial wave function profile can be
generated by bulk mass terms and the Yukawa couplings can be
determined by the wavefunction overlap of the Higgs and matter
fields. The bulk action for hypermultiplets $\{\Phi,\Phi^c\}$ with
mass terms is
\begin{eqnarray}
S_5=\int d^4x\int dy\[\int d^4\theta
\(\Phi^\da\Phi+(\Phi^c)(\Phi^c)^\da \)+\int d^2\theta {\Phi}^c
\(\partial_y+M_{\Phi}\){\Phi}\]~.~\,
\end{eqnarray}
In supersymmetric theories matter multiplets with kink bulk mass
terms still have zero modes. Depending on the sign of $M_{\Phi}$,
the zero mode is localized toward the $O$ or the $O'$ brane. The
zero mode wave function of $\Phi$ has a suppression factor
$\exp(-M_\Phi y)$ which means that the zero mode is localized near
$y=0$ for $M_\Phi>0$ and near $y=\pi R/2$ for $M_\Phi<0$. The
$M^{+-}$ (and $M^{-+}$) modes in the limit $M^{+-}\pi R/2\gg1$ (and
$M^{-+}\pi R/2\ll-1$) have the lightest KK mass
$M_{KK}=2|M_{zz^\pr}|\exp(-|M_{zz^\pr}|\pi R/2)$ which is less than
$1/R$.

We assume that the Yukawa couplings are localized on the $y=\pi R/2$
brane with the general form \beqa S=\int d^4x \int\limits_{0}^{\pi
R}d y\f{1}{2}\left[\delta\left(y-\f{\pi R}{2}\right)\pm\delta
\left(y+\f{\pi R}{2}\right)\right]\int d^2\theta
\f{y^{ijk}}{M_*^{3/2}}\Phi_i\Phi_j\Phi_k~, \eeqa where the Yukawa
couplings $y^{ijk}$ is assumed to be around ${\cal O}(4\pi)$, and
$M_*$ is the cutoff scale of the theory. This results in the four
dimensional Yukawa couplings
\beqa
W_{4D}=\la^{ijk}\phi_i\phi_j\phi_k~,
\eeqa
where
\beqa
\la^{ijk}\approx
\sqrt{Z[M(\phi_i)]Z[M(\phi_j)]Z[M(\phi_k)]}~y^{ijk}~,
\eeqa
with
\beqa \label{profile}
Z[M(\phi_i)]=\f{2M(\phi_i)}{M_*}\f{1}{e^{M(\phi_i)\pi R}-1}~.
\eeqa
Depending on the value of the bulk masses $M(\phi_i)$, we can have
different suppression factors for the Yukawa couplings. In this
paper, we assume that the Higgs fields $h$ and $\bar{h}$ are
strongly localized on the symmetry breaking $O^\pr$ brane which
implies $M_h,~M_{\bar{h}}\ll-1/R$.

Our goal is to explain the SM fermion masses and mixings based on
the deformed Froggatt-Nielsen mechanism via wave function profiles,
which is very difficult due to the flipping the right-handed up and
down type quarks. To solve this problem, we introduce an additional
discrete symmetry, and use the traditional Froggat-Nielsen mechanism
together with the wave function profiles to generate realistic SM
fermion masses and mixings. After embedding the matter multiplets in
flipped $SU(5)$, we can have three types of profiles: ${\bf
10}_{1}~(Q_L,D_L^c,\nu_L^c)$ type, $\bar{\bf 5}_{-3}~(U_L^c,L_L)$
type and the ${\bf 1}_{5}~(E_L^c)$ type. The relevant suppression
profiles can be realized through different bulk mass terms.

Realistic neutrino masses can be generated using the double see-saw
mechanism by introducing additional SM singlets $N_i$ which mix with
the ordinary neutrino sector. We can write the R-symmetry preserving
interaction terms for the singlets as~\footnote{It will become clear
later that the R-charge assignments are
$R(H)=R(\overline{H})=R(F_i)=R(\bar{f}_i)=R({l_i^c})=0$ while
$R(h)=R(\bar{h})=2$.}
\beqa
 W=y^s_{ab}\f{\psi_2}{M_*}F^a\overline{H}N^b+\f{1}{2}M_{ab} N_a N_b~,
\eeqa
where we introduced an additional unit R-charge field
$\psi_2$ which will also play a role in the SUSY breaking sector.
After $\psi_2$ and $N_H^c$ components of $\overline{H}$ acquire
VEVs, we can get the neutrino mass terms
 \beqa {\cal L}=y^u_{ab}(\nu_L)^a(\nu_L^c)^b v_u
  + y^s_{ab} (\nu_L^c)^a N^b v_R+\f{1}{2}M_{ab} N^a N^b~,
 \eeqa
 where $v_u=<\bar{h}>$, $v_R=v M/M_*$, and $M_{ab}\gg v_u$.

The neutrino mass matrix in the basis of $(\nu_L,\nu_L^c,N)$ is
\beqa
M_\nu=\left( \bea{ccc} 0&y^u v_u&0\\
(y^u v_u)^T&0&y^s v_R\\0&(y^s v_R)^T&M\eea \right)~.
\eeqa
 So we obtain the light Majorana neutrino masses as
 \beqa
 {\cal M}_\nu^M=(y^u v_u)\[(y^s v_R) M^{-1}(y^sv_R)^T\]^{-1}(y^u v_u)^T~.
\eeqa In the Frogatt-Nielsen mechanism, the Dirac neutrino mass
matrix is proportional to the product of matrices $F_i$ and
$\bar{f}_i$ describing the fermion profiles~\footnote{For
simplicity, we use the same notaion for the SM fermions and their
five-dimensional profiles.}
 \beqa {\cal M}^{\rm Dirac}_\nu \propto
 \left(\bea{c}\bar{f}_1\\\bar{f}_2\\\bar{f}_3\eea\right) \cdot
 \(\bea{ccc} F_1&F_2&F_3\eea \)~.
 \eeqa
 So the light neutrino mass matrix is
 \beqa
 {\cal M}_\nu^M& \propto
&\left(\bea{c}\bar{f}_1\\\bar{f}_2\\\bar{f}_3\eea\right)\cdot
\(\bea{ccc} F_1&F_2&F_3\eea \)(M_R)^{-1}\left(\bea{c}F_1\\F_2\\F_3
\eea\right)\cdot \(\bea{ccc} \bar{f}_1&\bar{f}_2&\bar{f}_3 \eea
\)~,\nn\\&\propto&\left(\bea{c}\bar{f}_1\\\bar{f}_2\\\bar{f}_3\eea\right)\cdot\(\bea{ccc}
\bar{f}_1&\bar{f}_2&\bar{f}_3\eea \)~.
\eeqa
 From the tri-bimaximal (or bi-maximal) mixings in the neutrino sector, we
can determine a possible ratio of the $\bar{f}_i$ profiles
\beqa
\bar{f}_1:\bar{f}_2:\bar{f}_3~\sim~1~:~1~:~1~.
\eeqa
 Thus, the neutrino mass matrix is proportional to
 \beqa
 M_\nu^M \propto \left(\bea{ccc}~1&~1&~1\\~1&~1&~1\\~1&~1&~1\eea\right)~,
 \eeqa
 and the unitary transformation matrix is
 \beqa
U_\nu^L\sim\left(\bea{ccc}~1&~1&~1\\~1&~1&~1\\~1&~1&~1\eea\right)~.
\eeqa
Using the following four-dimensional effective Yukawa terms
\beqa
 W=\f{\tilde{S}_i}{M_*}F_j\bar{f}_k\bar{h}
\eeqa
with SM singlet fields $\tilde{S}_i$ profiles
\beqa
 \f{\langle\tilde{S}_i\rangle}{M_*}~\sim~(~1,~1,~1)~,
\eeqa
we can obtain the ratios for the profiles of $F_i$
\beqa
F_1:~F_2:~F_3 \sim \la^8:~\la^4:~1
\eeqa
from the up-type quark mass ratio
\beqa
 m_t~:~m_c~:~m_u~\simeq ~1~:~\la^4~:~\la^8
\eeqa
and the $\bar{f}_i$ profiles.

 The reason to introduce
$\tilde{S}_i$ is to explain the bottom quark masses and quark CKM
mixings.  We consider the discrete symmetry $Z_3$ for $F_i$ in the
following, and then the above Yukawa couplings for up-type quarks
can be invariant under $Z_3$ by assigning suitable $Z_3$ quantum
numbers to $\tilde{S}_i$.

So the up-type quark mass matrix is
\beqa
 M_u \sim (M_\nu^{\rm Dirac})^T \propto
\left(\bea{ccc}~\la^{8}&~\la^8&~\la^8\\~\la^4&~\la^4&~\la^4\\~1&~1&~1
\eea\right)~.
\eeqa
 This up-type quark mass matrix leads to the unitary transformation matrix
\beqa
V_L^u\sim\left(\bea{ccc}~1&~\la^4&~\la^8\\~-\la^4 &~1&~\la^4\\
~\la^{8}&~-\la^4&~1 \eea\right)~,
\eeqa
 defined by $M_u^{\rm diag}=(V_L^u)^\da M_d (V_R^u)$.

From the $\bar{f}_i$ profiles and the charged lepton mass hierarchy
\beqa
(~m_\tau~:~m_\mu~:~m_e)~\simeq (~1~:~\la^2~:~\la^4)~,
\eeqa
 we can obtain the ratios of the $l_i^c$ profiles
 \beqa
(~l_1^c:~l_2^c:~l_3^c)=(~\la^4:~\la^2:~1)~.
\eeqa
 Thus, the charge lepton mass matrix is
 \beqa
 M_e \propto \left(\bea{ccc}~\la^{4}&~\la^2&~1\\~\la^4&~\la^2&~1\\~\la^4&~\la^2&~1
\eea\right)~.
 \eeqa
 The unitary transformation matrix for
$M_e=(U_L^e)^\da M_e^{\rm diag} V_R^e$ can be obtained via the
matrix $H=M_e M_e^\da$
\beqa
U_L^e=\left(\bea{ccc}~1&~1&~1\\~1&~1&~1\\~1&~1&~1 \eea\right)~.
\eeqa
Thus, the PMNS mixing matrix is given by
\beqa
 U_{PMNS}\sim (U_L^e)^\da U_\nu^L\sim
 \left(\bea{ccc}~1&~1&~1\\~1&~1&~1\\~1&~1&~1\eea\right)~,
\eeqa which can have tri-maximal (or bi-maximal)-like mixings. The
symmetric down-type quark mass matrix cannot be naively determined
from the $F_i$ profile ratios $(\la^8,\la^4,~1)$ to agree with the
observed mass hierarchy
 \beqa
m_b~:~m_s~:~m_d~\simeq ~1~:~\la^2~:~\la^4~.
 \eeqa
 In order to obtain the realistic down-type quark mass ratios and quark CKM
mixings, we introduce an additional discrete symmetry and use the
traditional Froggat-Nielsen mechanism. We consider an Abelian
$Z_3$ flavor symmetry with three one-dimensional representations:
a trivial representation $1$, and two others, $1^\pr(\omega)$ and
$1^{\pr\pr}(\omega^2)$ where $\omega^3=1$. The representation of
$F_i$ in terms of $Z_3$ is presented in Table~\ref{discrete}.
%%%%%%%%%%%%%%%%%%%%%%%%%%%%%%%%%%%%%%%%%%%%%%%%%%%%%%%%%%%%%%%%%
\begin{table}[htb]
\caption{The $Z_3$ quantum numbers for $F_i$ fields.}
\begin{center}
\begin{tabular}{|c|c|c|c|c|c|c|c|c|c|c|c|c|c|c|}
\hline & $F_1$&$F_2$&$F_3$&$\bar{f}_i$&$l_i^c$&$\tilde{S}_1$
&$\tilde{S}_2$&$\tilde{S}_3$&$S_1$&$S_2$&$S_3$&$S_{12}$&$S_{13}$&$S_{23}$
\\
\hline $Z_3$& 1 &$\omega$&$\omega^2$&1&1&1&$\omega^2$&$\omega $
&1&$\omega$&$\omega^2$&$\omega^2$&$\omega$&1 \\
\hline
\end{tabular}
\end{center}
\label{discrete}
\end{table}
%%%%%%%%%%%%%%%%%%%%%%%%%%%%%
The effective symmetric Yukawa terms for down-type quarks
are~\footnote{The following Yukawa terms are not the most general
ones consistent with the symmetry. However, we can introduce
additional discrete or $U(1)$ symmetries and assign suitable charges
to the SM fermions to forbid all the other extra terms.}
\small
\beqa W=y^u
h\[\f{S_1}{M_*}F_1F_1+\f{S_2}{M_*}F_2F_2+\f{S_3}{M_*}F_3F_3
+\f{S_{12}}{M_*}F_1F_2+\f{S_{13}}{M_*}F_1F_3+\f{S_{23}}{M_*}F_2F_3\]
\eeqa
\normalsize
With the suppression factors \beqa \f{<S_1>}{M_*}&\sim&
~1,~\f{<S_2>}{M_*}\sim \la^6,~\f{<S_3>}{M_*}\sim \la^{12}~,\nn\\
\f{<S_{12}>}{M_*}&\sim& ~\la^3,~\f{<S_{13}>}{M_*}\sim
\la^7,~\f{<S_{23}>}{M_*}\sim \la^{10}~, \eeqa we obtain the
following mass matrix for down-type quarks \beqa {\cal M}_d \propto
\left(\bea{ccc}~\la^{16}&~\la^{15}&~\la^{15}\\
~\la^{15}&~\la^{14}&~\la^{14}\\~\la^{15}&~\la^{14}&~\la^{12}
\eea\right)~, \label{Md-FSU5} \eeqa which leads to the unitary
transformation matrix in the down-type quark sector \beqa
V_{\rm L}^d\sim\left(\bea{ccc}~1&~\la&~\la^3\\~\la &~1&~\la^2\\
~\la^{3}&~\la^2&~1 \eea\right)~, \eeqa with $M_d^{\rm
diag}=(V_L^d)^\da M_d (V_L^d)$. The quark CKM mixing matrix is given
by \beqa
V_{CKM}=(V_L^u)^\da(V_L^d)\sim\left(\bea{ccc}~1&~\la&~\la^3\\~\la
&~1&~\la^2\\
~\la^{3}&~\la^2&~1 \eea\right)~, \eeqa which agrees with the
experimental data. We know that $m_b:m_t=\la^3:1$, so if we set
\beqa (~F_1,~F_2,~F_3)\sim(~\la^8,~\la^4,~1)~, \eeqa we can obtain
the profiles \beqa
(\bar{f}_1,\bar{f}_2,\bar{f}_3)&\sim& (~\la^9~,~\la^9~,~\la^9)~,\\
(~l_1^c,~l_2^c,~l_3^c)&\sim&(~\la^7,~\la^5,~\la^3)~. \eeqa Here we
set $m_t\sim \la^9$ and assume approximate $b-\tau$ unification
$m_b\sim m_\tau$. We also assume that there are appropriate
suppression factors for fields that contain $h$ and $\bar{h}$, and
then the total factor $\la^9$ may be absorbed in $h$ and $\bar{h}$
at low energy. From the orbifolding procedure we know that the
matter content in each generation arises from different boundary
conditions. Using to the profiles of $F_i,~\bar{f}_i$, and $l_i^c$,
we can easily obtain the bulk masses for various generations which
we will not give explicitly here.

Finally, we briefly present another scenario in which the observed
SM fermion masses and mixings can also be generated. We assume
\beqa
&& (~F_1,~F_2,~F_3) \sim(~\la^7,~\la^4,~1)~, \\
&& (~\bar{f}_1,\bar{f}_2,\bar{f}_3)\sim (~\la^{10}~,~\la^9~,~\la^9)~,\\
&& (~l_1^c,~l_2^c,~l_3^c)\sim(~\la^6,~\la^5,~\la^3)~,
\eeqa
and
\beqa
&& \f{<S_1>}{M_*}\sim \la^{2},~\f{<S_2>}{M_*}\sim
   \la^6,~\f{<S_3>}{M_*}\sim \la^{12}~,\nn\\
&& \f{<S_{12}>}{M_*} \sim
  \la^4,~\f{<S_{13}>}{M_*} \sim \la^8,~\f{<S_{23}>}{M_*}\sim \la^{10}~.
\eeqa
From this we obtain that the down-type quark mass matrix is
similar to that in Eq. (\ref{Md-FSU5}). The up-type quark mass
matrix, the charged lepton mass matrix and the neutrino mass
matrix are
\beqa
M_u \propto
\left(\bea{ccc}~\la^{8}&~\la^7&~\la^7\\
~\la^5&~\la^4&~\la^4\\~\la&~1&~1 \eea\right), ~~
M_e \propto
\left(\bea{ccc}~\la^{4}&~\la^3&~\la\\~\la^3&~\la^2&~1\\~\la^3&~\la^2&~1
\eea\right),~~
M_\nu^M \propto
\left(\bea{ccc}~\la^2&~\la&~\la\\~\la&~1&~1\\~\la&~1&~1\eea\right)
\eeqa

\section{Gauge Mediated Supersymmetry Breaking with Spontaneously R-symmetry Breaking}
\label{sec-3}
We know from the previous orbifolding procedure that
the five-dimensional $N=1$ SUSY, which is $N=2$ SUSY in four
dimensions, reduces to $N=1$ SUSY in four dimensions. We need to
break further the remaining $N=1$ SUSY and mediate the breaking
effects to the SM sector.

In general, the breaking of SUSY requires the presence of
R-symmetry~\cite{nseiberg}. However, an exact R-symmetry forbids
gaugino masses which is not acceptable. One possible solution is to
explicitly break the R-symmetry by introducing small R-symmetry
violation terms which leads to meta-stable vacua \cite{ISS,IS}. But
there is, in general, some tension between the acceptable gaugino
masses and sufficiently long-lifetime vacua. The other possibility
is to spontaneously break the R-symmetry in ${\rm
O}^{\prime}$Raifeartaigh models.

We know that the generalized ${\rm O}^{\prime}$Raifeartaigh model
can serve as the low energy description of dynamical SUSY breaking
in strongly coupled gauge theories. It is known that the tree-level
flat directions (pseudo-moduli) from local SUSY-breaking vacuum
always exist in the ${\rm O}^{\prime}$Raifeartaigh
framework~\cite{ray,kshih}. In most ${\rm O}^{\prime}$Raifeartaigh
models constructed before, the pseudo-moduli, which are charged
under R-symmetry, break the R-symmetry by acquiring VEVs through a
radiatively generated effective potential. It was shown
in~\cite{r02shih} that the necessary condition to break R-symmetry
at one loop via Coleman-Weinberg potential is the existence of a
field with R-charge $R\neq 0$ or 2, which is rather complicated to
evaluate in detail. It is however possible to spontaneously break
R-symmetry by the tree-level VEVs of the fields other than
pseudo-moduli~\cite{kshih,CDFM,sun1}. It is shown in~\cite{kshih}
that a theory of this type with direct gauge mediation leads to
vanishing gaugino masses at leading order in $F$. We want to use the
generalized ${\rm O}^{\prime}$Raifeartaigh model in the hidden
sector, with spontaneous R-symmetry breaking at tree level, to
generate non-vanishing leading order gaugino masses through indirect
gauge mediation.

We use a Carpenter-Dine-Festuccia-Mason (CDFM) like
model~\cite{CDFM,sun1,sun2} in the hidden sector to achieve
tree-level spontaneous R-symmetry breaking
\beqa
W=-fX+m
\psi_2\tl{\psi}_2+m\psi_3\tl{\psi}_3+\la_2 X
\psi_2\tl{\psi}_{3}+m_2\psi_2^2 ~.
\eeqa
The superpotential contains an R-symmetry
\beqa
R(X)=2~,~R(\psi_2)=-q(\tl{\psi}_3)=1~,~q(\tl{\psi}_2)=1~,~q(\psi_3)=3~.
\eeqa
The tree-level scalar potential is
\beqa
V&=&|-f+\la_2\psi_2\tl{\psi}_{3}|^2+|m \psi_2|^2+|m\tl{\psi}_3|^2\nn \\
&&
+|m\tl{\psi}_2+\la_2 X\tl{\psi}_{3}+2m_2\psi_2|^2
+ |m{\psi}_3+\la_2 X {\psi}_{2}|^2~.
\eeqa

We are interested in SUSY breaking without identically vanishing
$\psi_i$ and $\tl{\psi}_{i}$. We can require
$F_{{\psi}_2}=F_{\tl{\psi_3}}=0$ simultaneously by properly chosen
$\tilde{\psi}_2$ and $\psi_3$ with arbitrary $X$. The reduced
potential reads
\beqa
V=|-f+\la\psi_2\tl{\psi}_{3}|^2+|m \psi_2|^2+|m\tl{\psi}_3|^2~.
\eeqa
The minimum occurs at
\beqa
\psi_2\tl{\psi}_{3}=\f{\la f-m^2}{\la^2}~,~|\psi_2|=|\tl{\psi}_3|~,
\eeqa
for $\la f>m^2$. The non-zero VEVs can be parameterized as follows
\beqa
  \psi_2=r e^{i\theta}~,~
  \tl{\psi}_3=r e^{-i\theta}~,~ r=\sqrt{\f{\la f-m^2}{\la^2}}~,
\eeqa
with the R-Goldstone boson labeled by $\theta$. In this case
with non-vanishing $r$, the R-symmetry is broken everywhere in the
pseudo-moduli space.

SUSY breaking can be mediated to the visible sector via the
messengers $\phi_i$ and $\tl{\phi}_i$. We want to use the two gauge
singlets $\psi_2$ and $\tl{\psi}_2$ to couple to the messenger
sector directly. In the SUSY breaking hidden sector, $\psi_2$
develops non-zero VEV in its scalar component while $\tl{\psi_2}$
gets non-zero F-term. Their couplings to the messenger sector are
\beqa
W=\[\la_{ij}^\pr(\psi_2+\tl{\psi}_2)+m_{ij}\]\phi_i\phi_j
 ={\cal M}_{ij}\phi_i\phi_j~,
\eeqa
where $\phi_i$ and $\tl{\phi}_j$ are
messenger fields transforming in the $({\bf 5,-2})$ and $({\bf
\bar{5}, 2})$ representation of flipped $SU(5)$, respectively. We
can also introduce additional messengers in $({\bf 10,1})$ and
$({\bf \overline{10},-1})$ representations of flipped
SU(5)~\footnote{ If we introduce only the $({\bf 10,1})$ and $({\bf
\overline{10},-1})$ pair as messengers, the slepton masses will be
too small. It is advantageous to also introduce a $({\bf 5,-2})$ and
$({\bf \bar{5}, 2})$ pair. In four dimensions this can lead to
successful gauge coupling unification for flipped $SU(5)$ embedded
into $SO(10)$.}. We use the following form for ${\cal M}_{ij}$ with
$\det\la^\pr_{ij}\neq0$ and $\det m_{ij} =0$
\beqa
W=\la^\pr (\psi_2+\tl{\psi}_2)\sum\limits_{i}\phi_i\tl{\phi}_i
 +m^\pr\sum\limits_{i,j}\phi_i\tl{\phi}_j~,
\eeqa
with $R(\phi_i)+R(\tl{\phi}_j)=2$ in the second term.

The new terms do not spoil the original SUSY breaking vacuum. In
terms of the total superpotential, we have
\beqa
-F_{{\psi}_2}^*&=&\la_2 X \tl{\psi}_3+m
\tl{\psi}_2+2m_2\psi_2+\la^\pr\sum\limits_{i}\phi_i\tl{\phi}_i~, \\
-F_{\tl{\psi}_2}^*&=& m
{\psi}_2+\la^\pr\sum\limits_{i}\phi_i\tl{\phi}_i~.
\eeqa
With $\phi_i=\tl{\phi}_i=0$, the messenger sector will not spoil the SUSY
breaking vacua which have $F_{\tl{\psi}_2}^*\neq0$ and
$F_{{\psi}_2}^*= 0$.

In the case of tree-level spontaneous R-symmetry breaking, we
parameterize
\beqa
 \langle \psi_2+\tl{\psi}_2\rangle=M+\theta^2 F~,
 \eeqa
 with
 \beqa
 M=\sqrt{\f{\la f-m^2}{\la^2}}~, ~~ F=mM~.
\eeqa
We can use the wave function renormalization technique
proposed in \cite{giud-ratt} to calculate the gaugino masses and
squark masses if we require $m<<M$. Then the supersymmetry
breaking soft mass terms are
\beqa
M_r&=&\f{\al_r}{4\pi}\Lambda_G~,
~~~\Lambda_G=F \f{\pa}{\pa \psi_2}\log\det{\cal M}~,\\
m_{\tl{f}}^2&=&2 C_{\tl{f}}\(\f{\al_r}{4\pi}\)^2\Lambda_S^2~,
~~~\Lambda_S^2=\f{|F|^2}{2}\f{\pa^2}{\pa
\psi_2 \pa \psi_2^*}\sum\limits_{i=1}^N(\log |{\cal M}_i^2|)^2~.
\eeqa
In our case, the messengers couple to the SUSY breaking fields
which in general leads to the non-constant determinant
\beqa
&&\det{\[\la_{ij}^\pr(\psi_2+\tl{\psi}_2)+m_{ij}\]}=(\psi_2+\tl{\psi}_2)^n
G(m^\pr,\la^\pr)~,\\&&n=\sum\limits_{i=1}^N
\(2-R(\phi_i)-R(\tl{\phi}_i)\)~,
\eeqa
similarly to the case of
(extra)ordinary gauge mediation \cite{CFShin}\footnote{In our case
$R(\psi_2)=R(\tl{\psi}_2)=1$, so $R(\phi_i)+R(\phi_j)=1$ which is
slightly different from the formula in \cite{CFShin}.}. In our
messenger sector with $\det\la^\pr\neq0$, we have
\beqa
\det\[\la^\pr(\psi_2+\tl{\psi}_2) +m^\pr\]=(\psi_2+\tl{\psi}_2)^N\det \la^\pr~.
\eeqa
Thus, as we can see, the gaugino masses at leading order in $F$ are
non-vanishing.

On the other hand it is problematic to have a massless R-Goldstone
boson. Fortunately, such massless mode can became massive through
gravitational effects. For example, we can add a constant term $W_0$
to original superpotential $W_1$ to tune the cosmological constant
to zero (or to a tiny value). Such constant term will explicitly
break the R-symmetry, and then contribute to the R-axion mass. The
value of the constant $W_0$ in the total superpotential $W=W_0+W_1$
can be determined from the scalar potential in supergravity~\cite{rs}
\beqa
V(\phi^\da,\phi)&=&e^{K^2/M_{Pl}^2}
  \left[(K^{-1})^j_i\(W^i+\f{WK^i}{M_{Pl}^2}\)\(W_j^*+\f{W^*
K_j}{M_{Pl}^2}\)-3\f{|W|^2}{M_{Pl}^2}\right]
\eeqa
with the derivatives of the Kahler potential $K$ defined as
\beqa
K^i(\phi^\da,\phi)=\f{\pa K}{\pa \phi_i}~,
~~~~K^{i}_j=\f{\pa^2 K}{\pa\phi^{j\da}\phi_i}~.
\eeqa
A vanishing cosmological constant term in the scalar potential
requires $W_0$ to be
\beqa
F^2=3\f{W_0^2}{M_{Pl}^2}~.
\eeqa
Then the
axion acquires the following mass \cite{raxion}
\beqa
m^2_a= \f{8}{f_a^2}\f{W_0|<W_{1,i}K^{-1}_{,ij^*}K_j^*-3W_1>|}{M_{Pl}^2}\sim
\f{F^2}{M M_{Pl}}
\eeqa
where $f_a$ is the axion coupling
\beqa
f_a^2=\sum\limits_{ij}(v_iQ_i)(v_j^*Q_j)<K_{ij^*}>\sim M^2~.
\eeqa
Requiring the axion coupling $f_a$ to lie in the astrophysically and
cosmologically allowed window~\cite{kim}
\beqa
 0.5\times 10^9 ~{\rm GeV}<f_a\sim M<2.5\times 10^{12} ~{\rm GeV}~,
\eeqa
 we can estimate the SUSY breaking scale
 \beqa
 0.5\times 10^{14} {\rm (GeV)^2}\lesssim F\lesssim 2.5\times 10^{17}{\rm (GeV)^2}
 \eeqa
 with the requirement that the gaugino masses $\al_g F/(4\pi M)$ are at
the order of TeV. The axion mass is estimated to lie within $1 ~{\rm
GeV}$ to $1 ~{\rm TeV}$ which may be constrained by cosmological
effects similar to moduli fields~\cite{axioncon}. In our scenario,
the gravitino acquires a mass
 \beqa
 m_{3/2}\simeq\f{F}{\sqrt{3}M_{Pl}}
 \eeqa
 with order $10^{-5}~{\rm GeV}\lesssim M_{3/2}\lesssim 10^{-2}$ GeV and is the LSP.

\section{Gauge Coupling Unification}
\label{sec-4}
The bulk gauge symmetry $SO(10)$ is broken down to the
flipped $SU(5)$ on the $O^\pr$ brane by boundary conditions. We need
to break the remaining gauge symmetry further down to the SM gauge
group. This step is realized via the antisymmetric Higgs fields $H$
and $\overline{H}$. The Higgs fields can acquire VEVs through the
superpotential
\beqa W=Y(\overline{H}H-v^2) ~,~\,
\eeqa
where $Y$ is
a SM singlet field. To preserve SUSY, the F-term flatnesses for the
chiral fields $Y,~H$, and $\overline{H}$ give
\beqa F_Y&=&\overline{H}H-v^2=0~,\\
F_H&=&Y\overline{H}-8\pi \la H h=0~,\\
F_{\overline{H}}&=& YH-8\pi \bar{\la}\overline{H}\bar{h}=0~,
\eeqa
and then we have
\beqa
\f{H^2}{\overline{H}^2}=\f{\bar{\la}\bar{h}}{\la h}
=\f{\bar{\la}}{\la}\tan\beta~\sim~{\cal O}(1)~.~\,
\eeqa
So we can
anticipate that $\langle H \rangle \sim \langle \overline{H} \rangle
\sim v\equiv{M_{23}/g_{23}}$, where $g_{23}$ is the $SU(3)_C \times
SU(2)_L$ unified gauge coupling.

There are two possibilities for the mass scale $v$, which
characterizes the breaking of the flipped $SU(5)$. Large
GUT-breaking ($(g_5 v)^2>>M_C \equiv 1/R$) and small GUT-breaking
($(g_5 v)^2<<M_C$). Here $g_5$ is the five-dimensional coupling with
mass dimension $-1$. The large GUT-breaking
scenario~\cite{kr,biggut} greatly changes the mass spectra of the
gauge bosons that correspond to the broken generators of flipped
$SU(5)$. In this case there is no approximate flipped $SU(5)$
unification era for the orbifold zero modes. Thus, we are only
interested in the small GUT-breaking scenario in which the flipped
$SU(5)$ breaking effects in the brane are negligible. In this case
we have an approximate $\al_2$ and $\al_3$ unification era upon
$M_{23}$.

From the missing-partner mechanism, we know that the triplet
components of $h$ and $\bar{h}$ are much heavier than the doublet
components which will be considered as $H_d$ and $H_u$,
respectively. We assume that the mass scale for the $N_F$ pairs of
messengers ${\bf (5,-2)}$ and ${\bf (\bar{5},2)}$ (and for the $N_G$
pairs of ${\bf (10,1)}$ and ${\bf (\overline{10},-1)}$) is $M_E^2
\sim M^2 (>> F)$ and is determined by the R-axion constraints to lie
between $0.5\tm 10^9$ GeV and $2.5 \tm 10^{12}$ GeV.\footnote{It is
possible to split the triplets and doublets inside ${\bf (5,-2)}$
and ${\bf (\bar{5},2)}$ by the Yukawa couplings between the
messengers and $H$ and $\overline{H}$. However, such Yukawa
couplings can be forbidden by R-symmetry. So we simply prohibit
these Yukawa couplings by some discrete symmetries.} For simplicity,
we also assume that the Yukawa couplings among the messenger fields,
the SM fermions and Higgs fields are negligibly small.

In the small GUT-breaking scenario, the gauge couplings $\alpha_2$
and $\alpha_3$ unify into $SU(5)$ first. After that, $SU(5)$ unifies
with $U(1)_X$ into $SO(10)$. The RGE running of the gauge couplings
are
\beqa \f{d~\alpha_i}{d\ln E}=\f{b_i}{2\pi}\alpha_i^2~ \, ,
\eeqa
where $E$ is the energy scale and $b_i$ are the beta functions. The
running of the gauge couplings for $U(1)_Y,~SU(2)_L$, and $SU(3)_C$
are given by
\beqa
&&(b_1,b_2,b_3)=\(~\f{41}{10},-\f{19}{6},-7\)
~~~~{\rm
for}~ M_Z<E<M_S~,\\
&&(b_1,b_2,b_3)=\(~\f{33}{5},~1~,-3\) ~~~~{\rm for}~ M_S<E<M_{E}~,\\
&&(b_1,b_2,b_3)=\(~N_F+\f{12}{5}N_G+\f{33}{5},~N_F+3N_G+1,N_F+3N_G-3\)\nn\\
&&~~~~~~~~~~~~~~~~~~~~~~~~~~~~~~~~~~~~~~~~~~{\rm
for}~~M_E<E<M_{23}~.
\eeqa
The gauge coupling of $U(1)_Y$ is
normalized to the $SU(5)$ generator: $g_Y^2=\f{3}{5}g_5^2$. In the
messenger sector we introduce $N_F$ pairs of ${\bf (5,-2)}$ and
${\bf (\bar{5},2)}$ as well as $N_G$ pairs of ${\bf (10,1)}$ and
${\bf (\overline{10},-1)}$ multiplets.

The unification of $\alpha_2$ and $\alpha_3$ determines the
unification scale $M_{23}$ which is independent of $M_E$
\begin{eqnarray}
2\pi{[\alpha_2^{-1}(M_{Z})-\alpha_3^{-1}(M_{Z})]}
&=&\ln\[\(\f{M_S}{M_Z}\)^{\f{23}{6}}\(\f{M_{E}}{M_S}\)^4\(\f{M_{23}}{M_E}\)^4\]\nn\\
&=&4\ln\(\f{M_{23}}{M_S}\)+\f{23}{6}\ln\(\f{M_S}{M_Z}\)~.
\end{eqnarray}

After the unification of the $\al_2$ and $\al_3$ couplings, the
flipped $SU(5)\times U(1)_X$ gauge group will further unify into
$SO(10)$. The $U(1)_Y$ generator is the combination of $U(1)_X$ and
the diagonal generator of $SU(5)$. After the normalization of
$U(1)_Y$ to $SU(5)$, that is $\al_Y=5\al_{em}/(3\cos^2\theta_w)$,
the relation between the flipped $SU(5)$ gauge couplings and the
$U(1)_Y$ gauge coupling at $M_{23}$ can be obtained \beqa
\f{25}{\alpha_Y}=\f{1}{\al_5}+\f{24}{\al_X}~. \eeqa Here we
normalize the $U(1)_X$ gauge coupling $g_X Q_X$ so that the $Q_X$
charge has a factor ${1}/{\sqrt{40}}$ consistently with the
unification into $SO(10)$. As mentioned before, in orbifold models
with kink masses, the lightest KK modes can be as light as
$2M\exp(-M\pi R/2)$. We assume that the lightest KK mode is heavier
than $M_{23}$. The bulk matter multiplets of flipped $SU(5)$ at
$M_{23}$ will give (from the ${\bf 16}$ and ${\bf 16^\pr}$
representations of $SO(10)$) $N_G+1$ pairs of chiral fields in the
${\bf (10,1)}$ and ${\bf (\overline{10},-1)}$ representation
(including $N_G$ pairs of messengers); $N_F$ pairs of ${\bf (5,-2)}$
and ${\bf (\bar{5},2)}$ messenger multiplets (from ${\bf 10}$
representation of $SO(10)$)~\footnote{The pair of ${\bf (5,-2)}$ and
${\bf (\bar{5},2)}$ multiplets which lead to the MSSM Higgs $H_u$
and $H_d$ are localized Higgs fields.}; and $N_f=3$ families of
$({\bf (10,1)}, ~{\bf ({\bar 5},-3)},~{\bf (1,5)})$ multiplets (from
${\bf 16}$ representation of $SO(10)$) to account for the MSSM
matter content.

After integrating out contributions from all the KK modes the
one-loop gauge couplings have the form \cite{CKS}
\beqa
\f{1}{g_a^2(\mu)}=\(\f{1}{g_a^2}\)_{bare}
 +\f{1}{8\pi^2}\[\Delta_a+b_a\ln\f{M_*}{\mu}\]~,
\eeqa
where the cut off scale $M_*\simeq M_U$ is assumed to be large
enough compared to other mass parameters of the theory. Here $\mu$
is the scale below the lightest massive KK modes but higher than
$M_{23}$, $\Delta_a$ are threshold corrections due to massive KK
modes while $b_a$ are the 1-loop beta function due to zero modes.
The bare couplings here consist of several pieces~\cite{bare}
\beqa
\(\f{1}{g_a^2}\)_{bare}=\f{\pi
R}{2g_{5a}^2}+\f{\ga_a}{48\pi^3}M_*\pi R ~,~\,
\eeqa
where $\ga_a$
are the coefficients of UV-sensitive linearly divergent corrections.
In orbifold GUT which is strongly coupled at $M_*$, $g_{5a}^2$ and
$\ga_a$ are universal. So we have
\beqa
\(\f{1}{g_a^2}\)_{bare}=\f{1}{g^2_{GUT}}~.
\eeqa
The KK threshold correction $\Delta_a$ can be calculated for $SU(5)$ to be
\beqa
\Delta^{SU(5)}&=&5\ln(\f{M_*\pi R}{2})+\f{3}{2}\ln(Z^1_{10}Z^2_{10}Z^3_{10})
+\f{1}{2}\ln(Z^1_{\bar{5}}Z^2_{\bar{5}}Z^3_{\bar{5}})
+\f{1}{2}\ln(Z^1_{m}Z^2_{m}{\cdots}Z^{2N_F}_{m})\nn\\
&&+\f{3}{2}\ln(Z^1_{n}Z^2_{n}{\cdots}Z^{2N_G}_{n})
  +\pi R\sum\limits_{N_f=1}^3 \(M_{10^i}+M_{5^i}+M_{1^i}\)~\nn\\
&&+ \f{\pi R}{2}\sum\limits_{i=1}^{2N_F} M_{m^i}
+\f{\pi R}{2}\sum\limits_{i=1}^{2N_G} M_{n^i}~,
\eeqa
while for $U(1)_X$ they are
\beqa
\Delta^{U(1)_X}&=&\f{1}{4}\ln(Z^1_{10}Z^2_{10}Z^3_{10})
   +\f{9}{8}\ln(Z^1_{5}Z^2_{5}Z^3_{5})+\f{5}{8}\ln(Z^1_{1}Z^2_{1}Z^3_{1})
   +\f{1}{2}\ln(Z^1_{m}Z^2_{m}{\cdots}Z^{2N_F}_{m})\nn\\
&& +\f{1}{4}\ln(Z^1_{n}Z^2_{n}{\cdots}Z^{2N_G}_{n})
   +\pi R \sum\limits_{N_f=1}^3
   \(M_{10^i}+M_{5^i}+M_{1^i}\)~\nn\\
&&+ \f{\pi R}{2}\sum\limits_{i=1}^{2N_F}  M_{m^i}
  +\f{\pi R}{2}\sum\limits_{i=1}^{2N_G} M_{n^i}  ~.
\eeqa
Here $Z(M)$ is the profile suppression factor which appears in
Eq.~($\ref{profile}$). The various profiles can be deduced from the
hierarchy in Section~\ref{sec-2}.

\begin{figure}[htb]
\centering
\includegraphics[width=7cm]{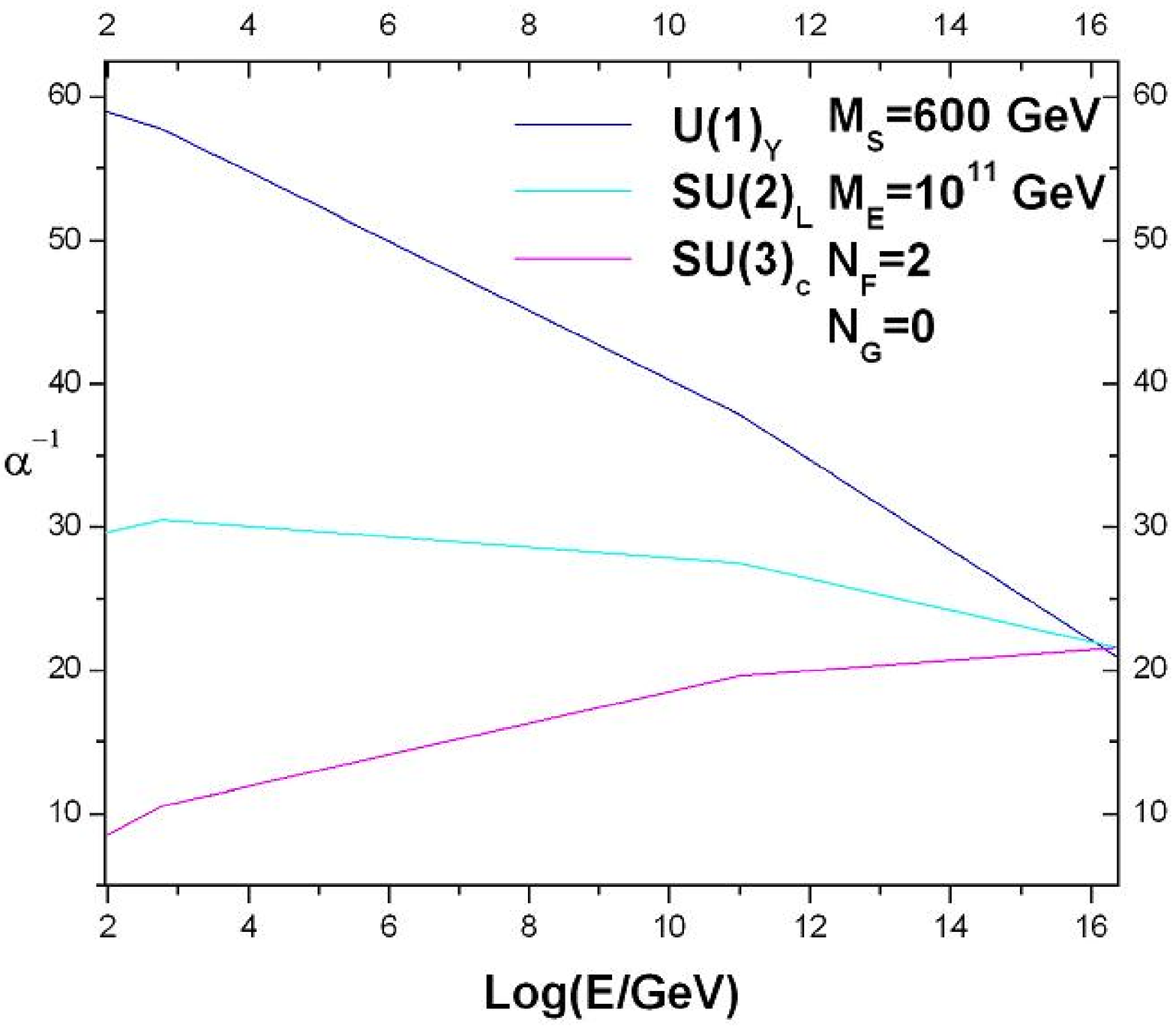}
\includegraphics[width=7cm]{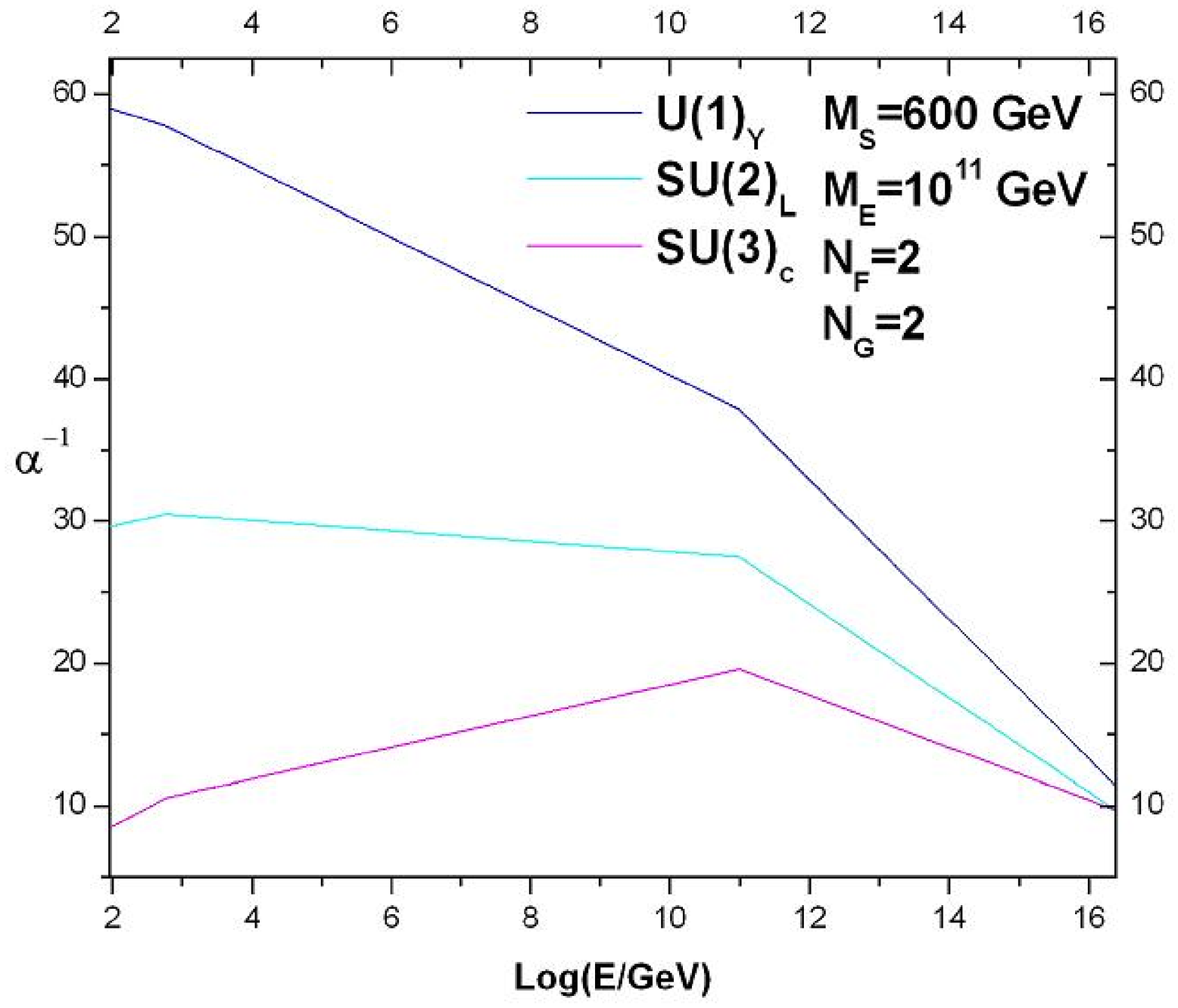}
\caption{One-loop RGE running of the three gauge couplings. The left
frame shows the case with $N_G=0$ while the right frame with
$N_G=2$. In the left frame, the unification of flipped $SU(5)$ into
$SO(10)$ is possible only when contributions of the profiles to the
threshold corrections are taken into account. } \label{fig1:FSU5}
\end{figure}

The zero mode contributions to the $SU(5)$ and $U(1)_X$ beta
functions above $M_{23}$ are calculated as
\beqa
(~b_5~,~b_X~)=(~N_F+3N_G-5~,~{N_F}+\f{1}{2}N_G+\f{15}{2})~.
\eeqa
Combining the previous expressions and the RGE running to $M_{23}$,
we can obtain in our model the relation of the gauge couplings at $M_{23}$
\beqa
2\pi\(\al_5^{-1}-\al_X^{-1}\)(M_{23})&=&
\(-\f{25}{2}+\f{5}{2}N_G\)\ln\(\f{M_*}{M_{23}}\)+5\ln(\f{M_*\pi R}{2})\nn\\
&&+ \f{5}{4}\ln(Z^1_{10}Z^2_{10}Z^3_{10})-\f{5}{8}\ln(Z^1_{5}Z^2_{5}Z^3_{5})
  - \f{5}{8}\ln(Z^1_{1}Z^2_{1}Z^3_{1})~\nn\\
&&+\f{5}{4}\ln(Z^1_{n}Z^2_{n}{\cdots}Z^{2N_G}_{n})~.
\eeqa
It is interesting to note that in our case when $N_G=0$ with $N_F$
messenger fields ${\bf (5,-2)}$ and ${\bf (\bar{5},2)}$, the cutoff
(strongly coupled unification) scale of the theory is independent of
the messenger profiles. Substituting the various profiles into the
above expression, we obtain
\beqa
2\pi\(\al_5^{-1}-\al_X^{-1}\)(M_{23})&=&-\f{25}{2}\ln\(\f{M_*}{M_{23}}\)
   +5\ln(\f{M_*\pi R}{2})\nn\\
&&+ \f{5}{4}\ln(\la^{12})-\f{5}{8}\ln(\la^{27})-\f{5}{8}\ln(\la^{15})\nn\\
&=&-\f{25}{2}\ln\(\f{M_*}{M_{23}}\)+5\ln(\f{M_*\pi R}{2})-11.25\ln\la\nn\\
&\simeq& -\f{25}{2}\ln\(\f{M_*}{M_{23}}\)+5\ln(\f{M_*\pi R}{2})+17.034
\eeqa
Our weak scale inputs \cite{PDG}
\beqa
M_Z&=&91.1876\pm0.0021 ~,~\,\\
\sin^2\theta_W(M_Z)&=&0.2312\pm 0.0002 ~,~\,\\
\alpha^{-1}_{em}(M_Z)&=&127.906\pm 0.019 ~,~\,\\
\alpha_3(M_z)&=&0.1187\pm 0.0020 \eeqa
fix the numerical values
of the standard $U(1)_Y$ and $SU(2)_L$ couplings at the weak scale
\beqa
\alpha_1(M_Z)&=&\f{5\alpha_{em}(M_Z)}{3\cos^2\theta_W}=(59.00048)^{-1}~,\\
\alpha_2(M_Z)&=&\f{\alpha_{em}(M_Z)}{\sin^2\theta_W}=(29.5718)^{-1}~.
\eeqa The unification scale $M_{23}$ can be determined after we set
the soft SUSY breaking mass scale $M_S$. For example, we can choose
$M_S=600~{\rm GeV}$ and obtain \beqa M_{23}=2.633\tm 10^{16} ~{\rm
GeV}~. \eeqa We present the RGE running of the various gauge
couplings below $M_{23}$ in Fig.~\ref{fig1:FSU5} for $N_G=0$ and
$N_G=2$, respectively.
%%%%%%%%%%%%%%%%%%%%%%%%%%%%%%%%%%%%%%%%%%%%%%%%%%%%%%%%%%%%%%%%%%%%%
%%%%%%%%%%%%%%%%%%%%%%%%%%%%%%%%%%%%%%%%%%%%%%%%%%%%%%%%%%%%%%%%%%%%%
%%%%%%%%%%%%%%%%%%%%%%%%%%%%%%%%%%%%%%%%%%%%%%%%%%%%%%%%%%%%%%%%%%%%%
%%%%%%%%%%%%%%%%%%%%%%%%%%%%%%%%%%%%%%%%%%%%%%%%%%%%%%%%%%%%%%%%%%%%%
In addition, we present the strongly coupled unification scales from
our numerical calculations for $N_G=0$ in Table~\ref{strong1-FSU5}.
These results are independent of the messenger scale $M_E$ and the
messenger numbers $N_F$. In this scenario with $N_F$ pairs of ${\bf
(5,-2)}$ and ${\bf (\bar{5},2)}$ messengers, the strongly coupled
unification is possible due to the threshold contributions of the
bulk matter profiles. The unification of flipped SU(5) into SO(10)
is not possible with such choice of messengers in four dimensions or
in orbifold models without kink mass terms.

If we adopt a nonzero $N_G$ and set the profile for ${\bf (10,1)}$
and ${\bf (\overline{10},-1)}$ to be ${\cal O}(1)$, we can get
\beqa
2\pi\(\al_5^{-1}-\al_X^{-1}\)(M_{23})&=&\f{5}{2}(N_G-5)\ln\(\f{M_*}{M_{23}}\)
 +5\ln(\f{M_*\pi R}{2})+\f{5}{4}\ln \la^{12}\nn\\
&&-\f{5}{8}\ln \la^{27}-\f{5}{8}\ln \la^{15}
  +\f{5}{4}\ln(Z^1_{n}Z^2_{n}{\cdots}Z^{2N_G}_{n})\nn\\
&\simeq&\f{5}{2}(N_G-5)\ln\(\f{M_*}{M_{23}}\)
 +5\ln(\f{M_*\pi R}{2})+17.034
\eeqa
with the last step obtained by taking
$Z_n^i=1$. The numerical results for strongly coupled unification
scale and non-zero $N_G$ are given in Table~\ref{strong2-FSU5}. In
fact, it is more advantageous to choose the case with $N_G\neq0$ not
only because it can realize successful unification in four
dimensions and ordinary orbifold models without kink mass terms, but
also because it can satisfy the consistency requirements that the
strongly coupled unification scale $M_U$ is much higher than
$M_C^*$.

%%%%%%%%%%%%%%%%%%%%%%%%%%%%%%%%%%%%%%%%%%%%%%%%%%%%%%%%%%%%%%%%%%%%%
%%%%%%%%%%%%%%%%%%%%%%%%%%%%%%%%%%%%%%%%%%%%%%%%%%%%%%%%%%%%%%%%%%%%%
\begin{table}[htb]
\caption{The strongly coupled unification scale $M_U=M_*$ versus the
compactification scale $M_C^*\equiv\pi R/2$ and the soft SUSY
breaking mass scale $M_S$ for $N_G=0$ in the units of GeV. The
symbol $``\backslash"$ marks the fact that no acceptable unification
occurs.}
\begin{center}
\begin{tabular}{|c|c|c|c|c|}
\hline $M_S[M_{23}]{\backslash} M_C^*$&$3.0\tm 10^{16}$&$4.0\tm
10^{16}$
&$5.0\tm10^{16}$&$6.0\tm 10^{16}$ \\
\hline $0.3\tm 10^3 [2.558\tm 10^{16}]$&$1.394\tm 10^{17}$&$1.151\tm
10^{17}$
&$9.918\tm 10^{16}$&$8.783\tm 10^{16}$\\
\hline $0.6\tm 10^3 [2.633\tm 10^{16}]$&$1.218\tm 10^{17}$&$1.006\tm
10^{17}$
&$8.668\tm 10^{16}$&$7.676\tm10^{16}$ \\
\hline $1.0\tm 10^3 [2.689\tm 10^{16}]$&$1.103\tm 10^{17}$&$9.107\tm
10^{16}$
&$7.848\tm 10^{16}$&$6.950\tm 10^{16}$\\
\hline $1.5\tm 10^3 [2.735\tm 10^{16}]$&$1.020\tm 10^{17}$&$8.416\tm
10^{16}$
&$7.253\tm 10^{16}$&$6.423\tm 10^{16}$\\
\hline $5.0\tm 10^3 [2.876\tm 10^{16}]$&$8.068\tm 10^{16}$&$6.660\tm
10^{16}$
&$5.740\tm 10^{16}$&$\backslash$\\
\hline
\end{tabular}
\end{center}
\label{strong1-FSU5}
\end{table}
%%%%%%%%%%%%%%%%%%%%%%%%%%%%%%%%%%%%%%%%%%%%%%%%%%%%%%%%%%%%%%%%%%%%%
%%%%%%%%%%%%%%%%%%%%%%%%%%%%%%%%%%%%%%%%%%%%%%%%%%%%%%%%%%%%%%%%%%%%%
\begin{table}[htb]
\caption{The strongly coupled unification scale $M_U=M_*$ versus the
compactification scale $M_C^*\equiv\pi R/2$ and the soft SUSY
breaking mass scale $M_S$ for $N_G=2$ in GeV units. The symbol
$``\backslash"$ signifies the fact that no acceptable unification
occurs.}
\begin{center}
\begin{tabular}{|c|c|c|c|c|}
\hline $M_S[M_{23}]{\backslash} M_C^*$&$8.0\tm 10^{16}$& $2.0\tm
10^{17}$
&$6.0\tm10^{17}$&$1.0\tm 10^{18}$ \\
\hline $0.3\tm 10^3 [2.558\tm 10^{16}]$&$2.946\tm 10^{20}$&$4.714\tm
10^{19}$
&$5.238\tm 10^{18}$&$1.886\tm 10^{18}$\\
\hline $0.6\tm 10^3 [2.633\tm 10^{16}]$&$1.883\tm 10^{20}$&$3.013\tm
10^{19}$
&$3.348\tm 10^{18}$&$1.205\tm10^{18}$ \\
\hline $1.0\tm 10^3 [2.689\tm 10^{16}]$&$1.354\tm 10^{20}$&$2.166\tm
10^{19}$
&$2.407\tm 10^{18}$&$\backslash$\\
\hline $1.5\tm 10^3 [2.735\tm 10^{16}]$&$1.042\tm 10^{20}$&$1.667\tm
10^{19}$
&$1.852\tm 10^{18}$&$\backslash$\\
\hline $5.0\tm 10^3 [2.876\tm 10^{16}]$&$4.788\tm 10^{19}$&$7.661\tm
10^{18}$
&$8.512\tm 10^{17}$&$\backslash$\\
\hline
\end{tabular}
\end{center}
 \label{strong2-FSU5}
\end{table}
%%%%%%%%%%%%%%%%%%%%%%%%%%%%%%%%%%%%%%%%%%%%%%%%%%%%%%%%%%%%%%%%%%%%%
%%%%%%%%%%%%%%%%%%%%%%%%%%%%%%%%%%%%%%%%%%%%%%%%%%%%%%%%%%%%%%%%%%%%%

\section{Proton Decay}
\label{sec-5} One of the unique GUT predictions is proton decay.
There are several sources in SUSY GUT models: (i) The conventional
lepto-quark vector gauge boson exchange which will lead to dimension
six baryon number violating operators; (ii) The new contributions
from supersymmetry.

The dominant new contribution in SUSY GUTs comes from the F-type
dimension five baryon number violating operators
\beqa
{\cal O}_{\Delta B\neq0}=\f{1}{M_U}Q_i^TC^{-1}\tau_2Q_j
\tilde{Q}_k^T\tau_2 \tilde{L}\epsilon^{ijk}~,
\eeqa
which can arise
from triplet Higgsino exchange in the presence of a triplet Higgsino
mass insertion term $M_H^T\tilde{\overline{H}}\tilde{H}$. Although
this operator cannot induce proton decay at the lowest order because
it is composed of squarks and sleptons, they can cause proton decay
once gaugino loops are included. Thus, we anticipate a proton
lifetime $\tau_P\sim (M_H^T)^2$ which may not be consistent with the
unification scale and then cause a problem. In the previous
discussions we pointed out that the D-T splitting problem in SUSY
GUTs is intimately related to the dimension five proton decay
problem. In flipped SU(5), the problem of D-T splitting can be
naturally solved via the elegant missing partner mechanism. In
particular, the mixing term between the triplet Higgsinos is absent
due to R-symmetry, thus it will not cause proton decay.

The direct $\mu$-term $\mu\bar{h}h$ is forbidden by the R-symmetry
because of the following reason. From the superpotential we have
\beqa R(H H
h)+R(\overline{H}\overline{H}\bar{h})=R(\bar{h}h)+2R(\overline{H}H)=4~.~\,
\eeqa The superpotential terms where $H$ and $\overline{H}$ acquire
VEVs indicate that $R(\bar{H}H)=0$ which means
$R(\bar{h}h)=4$.\footnote{We can set $R(\bar{h})=R(h)=2$ with
$R(\bar{H})=R(H)=0$. Thus all the matter multiplets in flipped
$SU(5)$ have a vanishing R-charge. } It is obvious that such
$\mu$-term is prohibited by R-symmetry. An effective $\mu$-term can
be generated through Giudice-Masiero mechanism~\cite{giudice-m} by
introducing some gauge singlets $Z$ with R-charge $4$. The effective
Kahler potential is \beqa K&=& \(\f{1}{\Lambda}Z^\da
h\bar{h}+h.c.\)+\f{1}{\Lambda^2}Z^\da Z h^\da
h+\f{1}{\Lambda^2}Z^\da Z\bar{h}^\da \bar{h}+\cdots ~,~ \eeqa while
the $B\mu$-term $Z^\dagger Z h\bar{h}/\Lambda^2$ is forbidden in the
potential. After the singlet $Z$ gets a VEV \beqa \langle Z \rangle
= Z_0 + \theta^2 Z_F \eeqa which breaks SUSY and R-symmetry, an
effective $\mu$-term can be generated: $\mu\sim Z_F/\Lambda$.
Although the $B\mu$-term is forbidden by R-symmetry, such term can
arise from gaugino loops and can be naturally small compared to the
$\mu$-term. The possible UV completion, which gives the interaction
between the singlet $Z$ and the hidden SUSY breaking sector, is
rather complicated. Thus, for simplicity we will not present a
realistic model here. The small effective $\mu$-term will not
reintroduce the proton decay problem since the decay process will
have an additional suppression factor $(\mu/M_H)^2$.

We can impose R-parity to forbid dimension-four proton decay
interactions. Additional interactions leading to dangerous dimension
five operators, besides those by heavy Higgsino exchange, can be
introduced on the gauge symmetry breaking $O^\pr$ brane as follows
\beqa W\sim [\delta(y-\pi R/2)+\delta(y+\pi R/2)]\f{(\psi_2)^2
}{M_{Pl}^{3}}\la^{abcd} F_a \bar{f}_b \bar{f}_c l^c_{d}~, \eeqa
after $\psi_2$ acquires a VEV. Here $a,~b,~c$, and $d$ are family
indices and the R-charge of the gauge singlets is $R(\psi_2)=1$. It
corresponds to an effective dimension-five operator suppressed by $
M_{pl}^3/M^2\sim 10^{30} {\rm GeV}$. Such operators will certainly
not violate the current proton decay lower bound.

\section{Conclusions}
\label{sec-6} We proposed a realistic flipped $SU(5)$ model from an
orbifolded $SO(10)$ model. The SM fermion masses and mixings were
obtained via the traditional Froggatt-Nielsen mechanism and the
five-dimensional wave function profiles of the SM fermions. The
breaking of $N=1$ supersymmetry after orbifolding was realized via
tree-level spontaneous R-symmetry breaking in the hidden sector and
extra(ordinary) gauge mediation. We generated realistic SUSY
breaking soft mass terms with non-vanishing gaugino masses. In
addition, we studied the gauge coupling unification in detail by
including the messenger fields at the intermediate scale and the KK
states at the compactification scale. We found that the $SO(10)$
unified gauge coupling is very strong and the unification scale can
be much higher than the compactificaiton scale. Finally, we briefly
commented on proton decay.

\section*{Acknowledgments}
This work was supported by the Australian Research Council under
project DP0877916, by the National Natural Science Foundation of
China under grant Nos. 10821504(TL), 10725526(JM) and
10635030(JM),by the DOE grant DE-FG03-95-Er-40917 (TL), and by the
Mitchell-Heep Chair in High Energy Physics (TL).

\end{document}